\documentclass[
reprint,prl,
amsmath,
twocolumn,
amssymb,aps,
bibliography=totoc]{revtex4-1}

\usepackage{dcolumn}% Align table columns on decimal point
\usepackage{bm}
\usepackage{appendix}
\usepackage{amsmath}
\usepackage{slashed}
\usepackage{multirow}
\usepackage{color}
\usepackage{mathtools}
\usepackage{graphicx}
\usepackage{braket}
\usepackage{setspace}
\usepackage{graphicx}
\usepackage{geometry}
\usepackage{hyperref}
\usepackage{starfont}
\usepackage{amssymb}
\usepackage{bbold}
\usepackage[shortlabels]{enumitem}
\usepackage{framed}
\usepackage{empheq}
\usepackage{amsmath}
\usepackage{tikz-feynman}
\usepackage{simpler-wick}
\usepackage{wasysym}
\usepackage{mhchem}
\usepackage{extarrows}

\DeclarePairedDelimiter\abs{\lvert}{\rvert}
\makeatletter
\let\oldabs\abs
\def\abs{\@ifstar{\oldabs}{\oldabs*}}
\makeatother

\def\bea{\begin{eqnarray}}
\def\eea{\end{eqnarray}}
\def\beq{\begin{equation}}
\def\eeq{\end{equation}}

\def\TeV{\,{\rm TeV}}

\def\sm{{\rm sm}}

\begin{document}

\title{Super heavy thermal dark matter}

\author{Hyungjin Kim}
\affiliation{%
Department of Particle Physics and Astrophysics, 
Weizmann Institute of Science, Rehovot 7610001, Israel}

\author{Eric Kuflik}
\affiliation{%
Racah Institute of Physics, 
Hebrew University of Jerusalem, Jerusalem 91904, Israel}

\begin{abstract}
We propose a mechanism of elementary thermal dark matter with mass up to $10^{14}$~GeV, within a standard cosmological history, whose relic abundance is determined solely by its interactions with the Standard Model, without violating the perturbative unitarity bound. The dark matter consists of many nearly degenerate particles which scatter with the Standard Model bath in a nearest-neighbor chain, and maintain chemical equilibrium with the Standard Model bath by in-equilibrium decays and inverse decays. The phenomenology includes super heavy elementary dark matter and heavy relics that decay at various epochs in the cosmological history, with implications for CMB, structure formation and cosmic ray experiments. 
\end{abstract}

\maketitle

\section{Introduction}

One of the biggest questions in fundamental physics is the nature of dark matter (DM). The possibility that
DM is a thermal Weakly Interacting Massive Particle (WIMP), whose abundance is determined by $2 \to 2$ annihilations into Standard Model (SM) bath particles, is exciting, but has alluded detection thus far. The WIMP is particularly intriguing  because it is very predictive---its  abundance is determined only by its interactions with the SM, which informs us how it may be detected. 

The WIMP paradigm has been a guide towards  the properties of DM, such as its mass and interactions. In particular, within the WIMP freezeout mechanism, there is a an upper bound on the DM mass from  perturbative unitarity, of  $ {\cal O}(100)\TeV$~\cite{Griest:1989wd}. 
The reason is that the WIMP annihilation rate is proportional to an exponentially decreasing DM density, and so the amount of dark matter that can be annihilated away before freezeout is limited by the theoretical size of the cross-section.

Of course, DM may be much heavier than this bound, if it is not a WIMP. 
Such models include non-thermal dynamics, decoupled dark sectors, inflationary and gravitational production, nonstandard cosmological histories, and large entropy production~\cite{Hui:1998dc,Kolb:1998ki,Chung:1998rq,Chung:2001cb,Feng:2008mu,Harigaya:2014waa,Davoudiasl:2015vba,Randall:2015xza,Dev:2016xcp,Harigaya:2016vda,Berlin:2016vnh,Berlin:2016gtr,Bramante:2017obj,Berlin:2017ife,Hamdan:2017psw,Cirelli:2018iax,Babichev:2018mtd,Hashiba:2018tbu,Hooper:2019gtx}. In none of these cases, however, is the DM   abundance solely determined by its interactions with the SM. The common lore is that an elementary DM candidate that is thermally coupled to the SM, within a standard cosmological history, cannot have mass well above the WIMP perturbative unitarity bound. Exceptions include composite DM, but still do not go much beyond the above bound~\cite{Harigaya:2016nlg,Geller:2018biy,Smirnov:2019ngs}. 
 
In this {\it Letter} we present a new freezeout mechanism within a standard cosmological history.  The DM is an elementary particle that is thermalized with the SM at high temperatures, its relic abundance is determined via its freezeout from the SM bath, and the DM mass can be as high as $10^{14}$ GeV for s-wave processes,  without violating the perturbative unitarity limit. In future work, we show how Planck-scale DM can be reached for velocity-dependent processes~\cite{future1}. 
 
 The general idea is as follows. The dark matter consists of $N$ approximately degenerate states, $\chi_i$ ($i=1..N$). These states co-scatter~\cite{DAgnolo:2017dbv} off of the SM bath, but only in a chain of nearest-neighbor interactions
 \beq
\chi_i + \sm \leftrightarrow \chi_{i+1}+ \sm,
\label{chainintro}
\eeq
while the $N^{\rm th}$ state co-decays~\cite{Bandyopadhyay:2011qm,Dror:2016rxc,Kopp:2016yji,Okawa:2016wrr,Dror:2017gjq,Dery:2019jwf} in equilibrium with the SM,
\beq
\chi_N \to \sm + \sm\,.
\label{codecay}
\eeq
Here $\chi_1$ is the DM candidate. This setup is summarized in Fig.~\ref{figchain}.  The processes in  Eq.~\eqref{chainintro} can maintain  chemical equilibrium much longer than annihilations can, because the interaction rate for the scattering
$
\Gamma_{\rm sctr} = n_{\rm SM} \left< \sigma v \right>_{\rm sctr}  \label{eq:sctr}
$
 never becomes exponentially suppressed.  The in-equilibrium decay allows for the whole system to have vanishing chemical potential for a long time---if the $N^{\rm th}$ particle was annihilating with the bath, the system would inherit the unitarity bound from co-annihilations. Finally, the chain, which will typically require $N>5-20$, depending on the DM mass,  ensures the stability of the DM $\chi_1$.

\section{General Idea}

Consider a DM particle $\chi_1$, whose density changes via scattering with a light SM bath particle,
\beq
\chi_1 + \sm \leftrightarrow \chi_2+ \sm,
\label{chain2}
\eeq
where $\chi_2$ has similar mass to $\chi_1$.
This process can maintain chemical equilibrium much longer than annihilations with the same interaction strength, because the interaction rate for the scattering does not depend on the DM density, and therefore the rate does not become exponentially suppressed.  
The DM is able to maintain equilibrium to smaller temperatures, becoming more Boltzmann suppressed than a WIMP for the same size cross-section.

However, $\chi_1 $ can only reduce exponentially for as long as  $\chi_2$ reduces exponentially (for instance, by maintaining chemical equilibrium with the SM bath). 
 Thus in order to go beyond the unitarity bound on annihilations, some other process is still needed to reduce the $\chi_2$ density. If this process is annihilations (such as $ \chi_2+ \chi_2 \leftrightarrow~\sm + \sm $), then it will freezeout too early, and the unitarity bound on annihilations will apply again. If instead, $\chi_2$ decays equilibrium with the SM bath, chemical equilibrium can be maintained for much longer. Thus, our proposed mechanism is a combination of co-decay and co-scattering dark matter. 
 
\begin{figure}
\[
\begin{tikzpicture}[baseline=(current bounding box.center)]
  \begin{feynman}[small,  baseline=(a)]
    \node[blob] (a);
    \vertex [above left =of a] (i1) {\(\chi_1\)};
    \vertex [below left=of a] (i2)   {\(\rm sm\)};
    \vertex [above right =of a] (f1) {\(\chi_2\)};
    \vertex [below right=of a] (f2)  {\(\rm sm\)};
    \diagram* {
    (i1) -- (a)-- (f1),
    (i2) -- (a) -- (f2), 
    };
  \end{feynman}
\end{tikzpicture}
~~
\begin{tikzpicture}[baseline=(current bounding box.center)]
  \begin{feynman}[small,  baseline=(a)]
    \node[blob] (a);
    \vertex [above left =of a] (i1) {\(\chi_2\)};
    \vertex [below left=of a] (i2)   {\(\rm sm\)};
    \vertex [above right =of a] (f1) {\(\chi_3\)};
    \vertex [below right=of a] (f2)  {\(\rm sm\)};
    \diagram* {
    (i1) -- (a)-- (f1),
    (i2) -- (a) -- (f2), 
    };
  \end{feynman}
\end{tikzpicture}
\cdots
\begin{tikzpicture}[baseline=(current bounding box.center)]
  \begin{feynman}[small,  baseline=(a)]
    \node[blob] (a);
    \vertex [above left =of a] (i1) {\(\chi_{_{N-1}}\)};
    \vertex [below left=of a] (i2)   {\(\rm sm\)};
    \vertex [above right =of a] (f1) {\(\chi_{_N}\)};
    \vertex [below right=of a] (f2)  {\(\rm sm\)};
    \vertex [right=.8cm of f1] (c) ;
    \node [below=.2cm of c, dot] (b) ;
      \vertex [ right=.75cm of b] (f3) {\(\rm sm\)};
       \vertex [ below=.55cm of b] (d) ;
         \vertex [ right=.2cm of d] (f4) {\(\rm sm\)};
    \diagram* {
    (i1) -- (a)-- (f1) -- (b) -- (f3),
    (i2) -- (a) -- (f2), 
    (b) -- (f4),
    };
  \end{feynman}
\end{tikzpicture}
\]
\caption{\label{figchain} Freezeout mechanism: the dark matter consists of many nearly degenerate particles which scatter with the Standard Model bath in a nearest-neighbor chain, and maintain chemical equilibrium with the Standard Model bath by in-equilibrium decays and inverse decays.}
\end{figure}
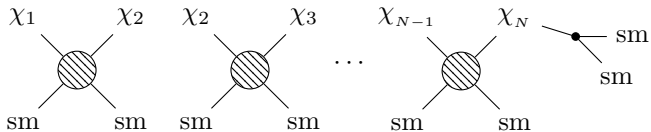

 However, one can easily see that the combination of the scattering process Eq.~\eqref{chain2}, and the in-equilibrium decay process  $\chi_2 \to \sm + \sm$, will necessarily lead to the fast decay of  $\chi_1$ via an off-shell $\chi_2$. In the presence of a  co-scattering chain,  such that scatters take place only for nearest neighbors
 \beq
\chi_i + \sm \leftrightarrow \chi_{i+1}+ \sm,
\label{chain}
\eeq
which codecay in equilibrium with $\chi_N$,
\beq
\chi_N \to \sm + \sm\,.
\eeq
$\chi_1$ can instead be long lived; 
 here $\chi_1$ is still the DM, but its decay width is suppressed due to the large phase space needed to decay to the SM ($\chi_1$ decays to  $2N$ SM particles). This is summarized in Fig.~\ref{figchain}.

\section{Two particle case}

As a toy example, we first work through the simplest case of $N=2$ dark matter particles with mass $m$. While this example is not cosmologically viable, it demonstrates the basic idea of the mechanism.  
Consider two degenerate states $\chi_1$ and $\chi_2$, which co-scatter~\cite{DAgnolo:2017dbv} off  of SM bath particles via the process 
\beq
\chi_1+ \sm \leftrightarrow \chi_2+ \sm.
\eeq
 The dark matter candidate is  $\chi_1$, while $\chi_2$ decays in equilibrium with the SM bath. 

The Boltzmann equations for the system are
\bea
\dot{n}_1 + 3 H n_1 &=& n_\sm \langle \sigma v \rangle (n_2 -n_1) ,
\nonumber\\
\dot{n}_2+ 3 H n_2 &=& n_\sm \langle \sigma v \rangle (n_1 -n_2) - (n_2 - n_2^{\rm eq}) \Gamma_{2},
\eea
where $\langle \sigma v \rangle$ is a thermally averaged cross section for the scattering process, $n_{\rm eq}$ is the equilibrium number density of $\chi$, and $\Gamma_2$ is the decay rate of $\chi_2$ in the thermal bath.  
This system is similar to the  co-scattering scenario~\cite{DAgnolo:2017dbv}, but here the DM bath is kept in equilibrium with the SM bath via decays and inverse-decays, rather than annihilations. Ultimately, since decays  and inverse-decays can stay in equilibrium much longer, this will allow for much heavier DM.

Unlike the case for freezeout via annihilations, the instantaneous freezeout approximation will not give a good estimate of the relic abundance. This is because the rate for $\chi_1$ scattering, $\Gamma \sim  n_\sm \langle \sigma v\rangle $, is not dropping off exponentially fast with the expansion, and therefore freezeout takes a long time. However, an approximate analytic solution to the relic abundance can still be determined from the Boltzmann equations, as we now detail.

If the decay rate  $\Gamma_2$ is larger than the Hubble expansion parameter when the temperature $T$ of the universe is equal to the DM mass, the number density of $\chi_2$ closely follows its equilibrium value. An approximate solution to the $\chi_1$ density can then be found by considering the single equation
\bea
Y_1' = \frac{\lambda}{x^2}(-Y_1 + Y_{\rm eq}),
\label{n1}
\eea
where $Y_i = n_i /s$ with $s$ the entropy density, $x=m/T$ and $\lambda = (n_\sm \langle \sigma v \rangle / H)|_{x=1}$. We have also assumed that the thermally averaged cross-section is velocity independent and took the number of relativistic degrees of freedom $g_\star=g_{\star s}$ to be constant.  
The asymptotic value of the relic abundance is
\bea
Y_1(\infty) \approx \frac{45 g_\chi}{2^{3/2} \pi^3 g_{\star s}} \lambda e^{-2\sqrt{\lambda}} \equiv Y_{\infty}(\lambda)\,,
\label{relic_N2}
\eea
where $g_\chi$ is the number of internal degrees of freedom of a $\chi$ particle. 

From the Boltzmann equation \eqref{n1}, $\chi_1$ departs equilibrium when ${\lambda}/{x_{\rm fo}^2} \sim 1$.
The fact that the relic abundance then scales as $Y_1(\infty) \sim e^{- 2 x_{\rm fo}}$ and not as $e^{- x_{\rm fo}}$ (as one finds in the instantaneous freezeout approximation), is the result of the slow freezeout.  At this point $\chi_1$ creation stops, but $\chi_1$ can  continue to scatter away. Neglecting the inverse process, solving
\beq
Y_1' = - \frac{\lambda}{x^2}Y_1 ,
\eeq
yields that $Y_1(x) = Y_1(x_{\rm fo}) e^{\frac{\lambda}{x} - \frac{\lambda}{x_{\rm fo}}} {\to}\, e^{- 2 x_{\rm fo}}$. One also sees from this solution that the abundance stops changing significantly when $x_{\rm fin}\sim \lambda = x_{\rm fo}^2$.

From the above estimation, one can  find $\langle \sigma v \rangle$ that reproduces the observed relic abundance of dark matter, while satisfying the unitarity bound. For instance, parameterizing the cross-section as $\langle \sigma v \rangle \equiv \alpha^2 / m_\chi^2$, one finds that for $\alpha \simeq 1$, ${m_\chi = 6 \times 10^{14}}$ GeV reproduces the correct relic abundance (assuming the DM scatters off of 4  SM degrees of freedom and that $g_{\star s}= 106.75$). 

\section{ $N>2$ Degenerate Case}
\begin{figure}[t]
\centering
\includegraphics[scale=0.43]{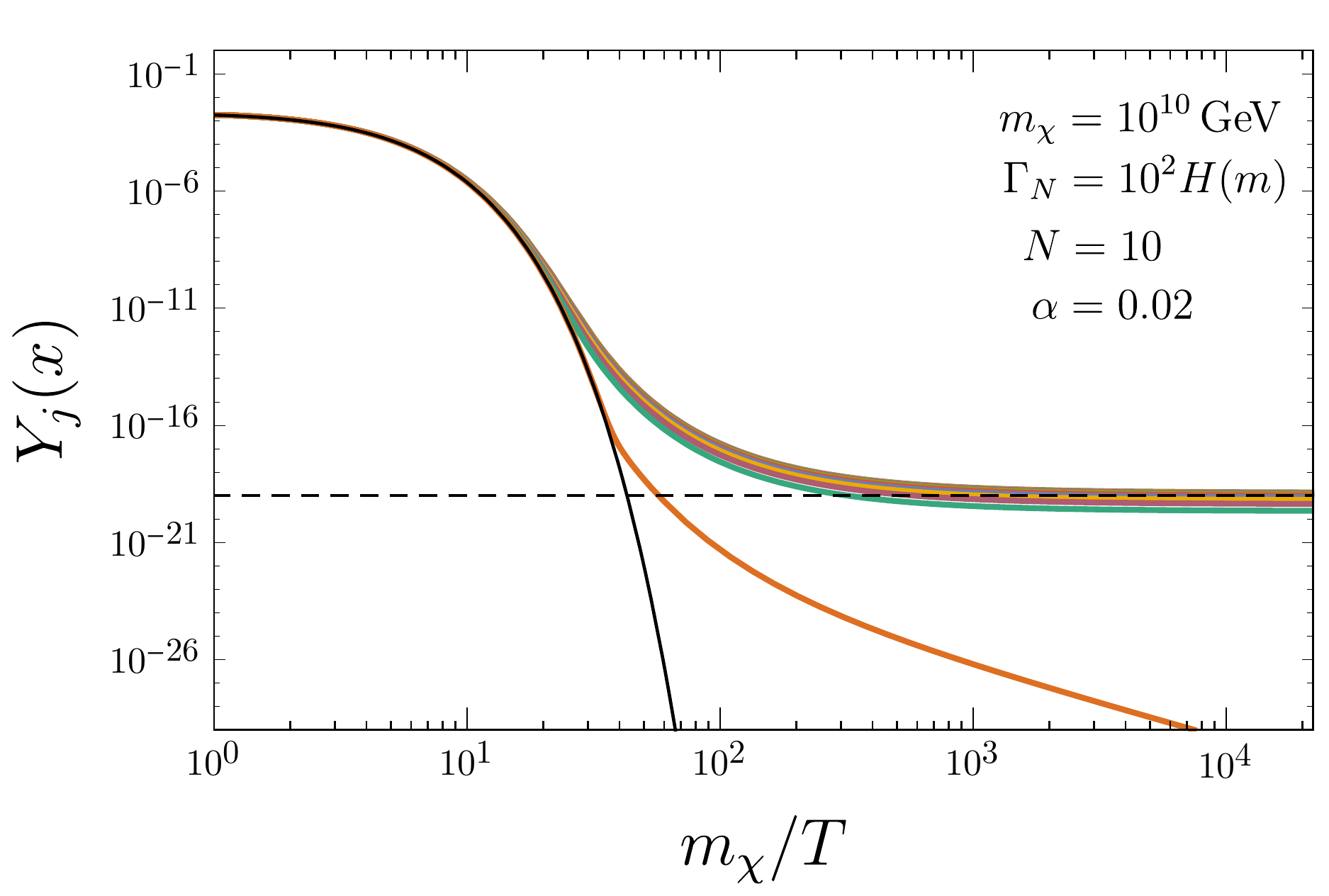}
\caption{Relic abundance of each particle with temperature evolution. 
Here the masses are assumed to be degenerate, and the decay rate of $\chi_N$ is chosen as $\Gamma_N = 10^2 H(m)$ to ensure that it decays in-equilibrium.
The horizontal dashed line is the observed relic abundance of DM, while the solid black curve is the equilibrium abundance.}
\label{fig:Yield}
\end{figure}

Although the simplest $N=2$ example does not work, it indicates a clear path forward to a viable setup: suppressing the decay rate of $\chi_1$.  
We thus consider $N >2 $ particles with the same type of nearest-neighbor interactions as before.
Similarly, we assume that only $\chi_N$ is able to decay into SM particles, that the masses are degenerate, and that the cross section for each interaction is the same. 
The equations of the system are
\bea
Y'_1 &=& \frac{\lambda}{x^2}  ( - Y_1 + Y_2),
\label{first_eq}
\nonumber\\
Y'_j &=& \frac{\lambda}{x^2} ( Y_{j-1} -2 Y_{j} + Y_{j+1} ) ,
\nonumber\\ 
Y_N' &=& \frac{\lambda}{x^2}  ( Y_{N-1} - Y_N) - x \lambda_d (Y_N - Y_N^{\rm eq}) ,
\label{last_eq}
\eea
where the second equation is valid for $j=2,\cdots,N-1$, $\lambda = (n_\sm \langle \sigma v \rangle / H)|_{x=1}$, and $\lambda_d = ( \Gamma_N / H)_{x=1}$.
Here prime denotes a derivative with respect to $x$. 
We also ignore the time variation of $g_\star$ and $g_{\star s}$, and assume that the cross section is constant in the non-relativistic limit. 
Note also that we are interested in parameter space where $\lambda,\,\lambda_d \gg 1$. 

This system can be solved in similar fashion to the $N=2$ case, by assuming $Y_N = Y_{\rm eq}$ and diagonalizing the differential equations. However, we can more simply obtain a solution with the new $N$ scaling by taking the large $N$ limit, and taking flavor space to be continuous. For $N\gg1$, the system behaves as a random walk between $\chi_j$ states, with a reflecting wall at $j=1$ and an absorbing cliff at $j=N$ (corresponding to the decay). Taking the indices of $j$ as a continuous variable $\ell = \pi j / [2 (N-1)]$, the system is described by the diffusion equation 
\bea
(\partial_\tau - D \partial_\ell^2) Y_\ell(\tau) = 0,
\eea
where $\tau = - 1/x$ and $D = \pi^2 \lambda/ [4 (N-1)^2]$ is a diffusion coefficient, and the boundary conditions are
\bea
\partial_\ell Y |_{\ell = 0} = 0,
\qquad
Y_{\pi/2} (\tau) = Y^{\rm eq}(\tau) 
\eea

From the form of the diffusion equation, we see that the system depends on the cross-section only through the combination $D \simeq \pi^2 \lambda/ 4 N^2$. From solving the boundary conditions,  the asymptotic profile ${Y_\ell (\infty) \propto \cos \ell}$. The solution of the diffusion equations gives 
the relic abundance
\beq
Y_i(\infty) \simeq 
Y_{\infty}(D) \frac{4}{\pi} \cos \Big( \frac{\pi}{2} \frac{j}{N} \Big)\,,
\label{relicY}
\eeq
where $Y_{\infty}$ is defined in Eq.~\eqref{relic_N2}.

In Fig.~\ref{fig:Yield} we plot the yield of the $\chi_i$ particles for $N=10$, solving the full Boltzmann equations, Eqs.~\eqref{last_eq}. We find that the relic abundances match the analytical estimate of Eq.~\eqref{relicY}. 

Now, one must check that $\chi_1$  is sufficiently long lived to be a viable DM candidate. A model-independent bound on the DM lifetime is $\tau > 5 \times 10^{18}$~sec~\cite{Audren:2014bca} (for earlier studies, see Refs.~\cite{Ichiki:2004vi,PalomaresRuiz:2007ry,Lattanzi:2007ux,Lattanzi:2008ds,Gong:2008gi,DeLopeAmigo:2009dc,Peter:2010au,Peter:2010sz,Wang:2010ma,Peter:2010jy,Huo:2011nz,Aoyama:2011ba}). However, if the DM decays to SM particles,  constraints on the lifetime may be much stronger,  $\tau >  10^{27}$~sec~(see e.g. Refs.~\cite{Cirelli:2012ut,Essig:2013goa,Blanco:2018esa}).

The decay of $\chi_1$ to SM particles takes place only through $N-1$ off-shell $\chi_i$ particles, into $2N$ SM particles, with decay width 
\bea
\Gamma_1 = \frac{1}{2m}  \int d\Phi_{2N} |{\cal M}|^2\,.
\eea
Treating the SM particles as massless, the total $2N$-body phase space is~\cite{Kleiss:1985gy}
\bea
\int d\Phi_{2N} = \frac{1}{S}\frac{2\pi}{\Gamma(2N) \Gamma(2N-1)} \frac{m^{4(N-1)}}{(16\pi^2)^{2N-1}}\,  ,
\eea
where $S$ is a symmetry factor accounting for identical particles in the final state.  
We may approximate the squared matrix element of the decay as~\footnote{We approximate internal propagators as $1/m^4$. One might wonder if the largest contribution arises from internal propagators being almost on-shell. Such possibility arises only with a small phase space density, so the parametric dependence of the decay rate on $m$ does not change. Nevertheless, internal propagators generally lead to larger $\Gamma_1$, but this changes our estimate on $N$ only by $\Delta N \sim 1\textrm{--}2$ for the range of $N$ of our interest.  }
\bea
|{\cal M}|^2 \simeq S^2 \frac{1}{m^{4N-4}} |{\cal M}|_{\chi \sm \to \chi \sm}^{2(N-1)} |{\cal M}|^2_{\chi_N \to \sm+ \sm}
\eea
where the factor $S^2$ accounts for expected number of diagrams. 
The decay rate of $\chi_1$ is then estimated as
\bea
\frac{\Gamma_1}{\Gamma_N} \simeq \frac{S}{(2N-1)!(2N-2)!} \left( \frac{\alpha^2}{16\pi^3} \right)^{N-1}\,. 
\label{decaychain}
\eea

\begin{figure}[t!]
\centering
\includegraphics[scale=0.4]{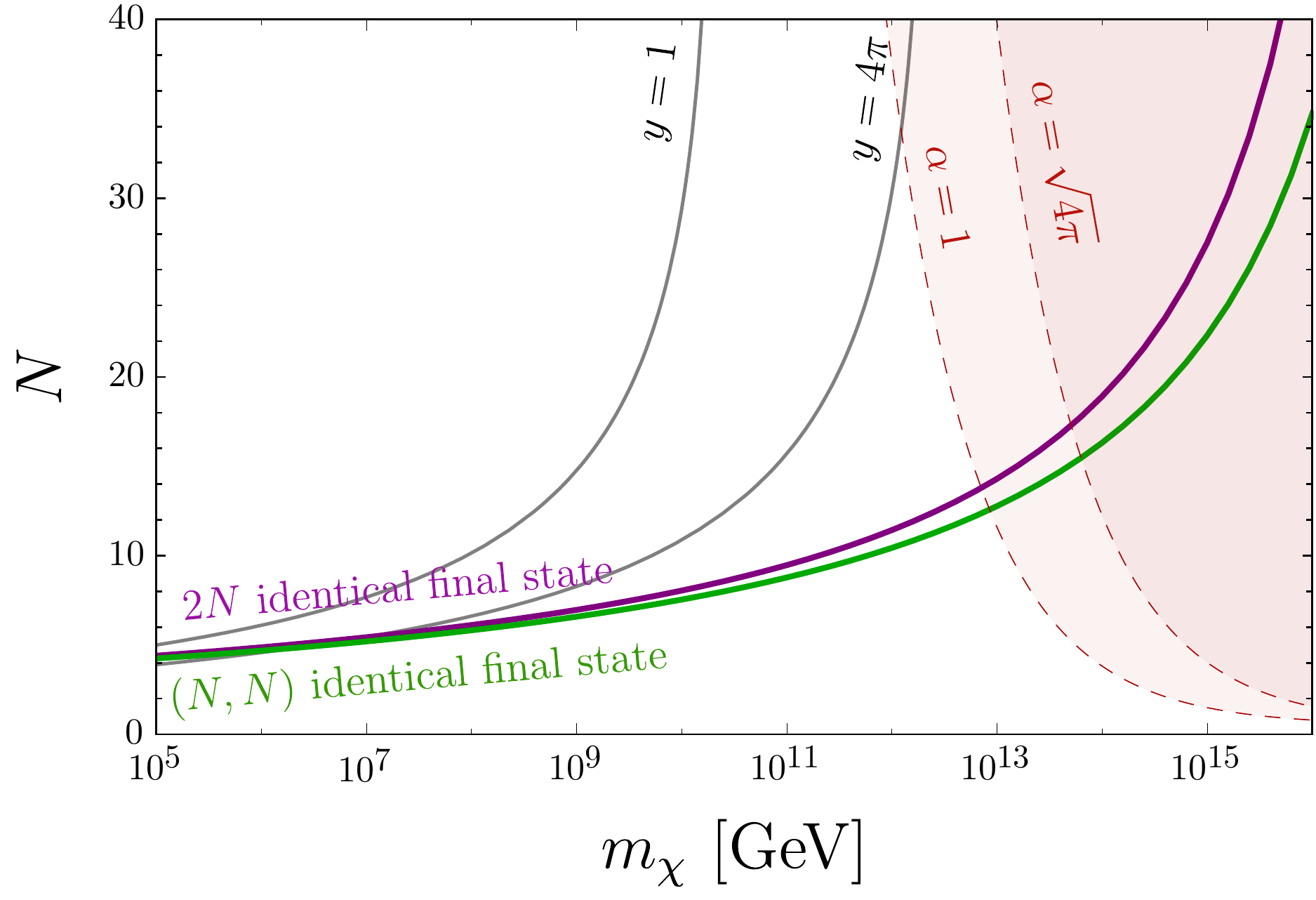}
\caption{Number of dark matter particles versus DM mass. 
The colored solid lines depict the minimal number of particles when $\chi_1$ decays through the chain~(see Eq.~\ref{decaychain}), while the gray solid lines depict the same within an explicit model~(see Eq.~\ref{decaymixing}). 
For both, we require that $\tau_1 > 10^{27}$~sec. 
The purple line corresponds to $\chi_1$ decay into $2N$ identical SM particles, while the green line corresponds to a pair of $N$ identical SM particles in the final state.
The dashed lines are contours of constant $\alpha$ that reproduce the observed relic abundance.}
\label{fig:minima_N}
\end{figure}

Assuming the $\chi_N$ decay is in equilibrium $\Gamma_N > H|_{x=1}$, requiring the stability of the DM, $\tau >  10^{27}$ sec, places a lower bound on the number of dark matter particles $N$. For instance, assuming that the DM decays to $2N$ identical SM particles corresponding to $S=(2N)!$, we find for $m= 10^{7}$~GeV that $N\ge 5$, and for $m= 10^{13}$~GeV that $N\ge14$. In Fig.~\ref{fig:minima_N}, we plot the minimum value of $N$ as a function of $m_\chi$, satisfying the lifetime requirement.

\section{Non-degenerate masses and couplings}
Up until now, we have considered the case of exactly degenerate masses, but it is natural to consider  small mass splittings between the particles. This  will have two important effects: creating forbidden channels and allowing the heavier states to decay directly into the lighter states.

In  standard two-particle forbidden annihilations, the mass splitting is negligible if it is smaller than the temperature when the abundance is still evolving, {\it i.e.}, $\delta m/m < x_{\rm fo}^{-1}$. For the chain interactions, a similar condition is found:
\beq
 (m_{\rm max} - m_{\rm min})/m_{\rm min}  < x_{\rm fin}^{-1} =  (\pi^2 \lambda/ 4 N^2)^{-1},
 \eeq 
 where $m_{\rm max}  $ and $m_{\rm min} $ is the maximum and minimum particle mass in the chain.
 
 If the mass splitting is larger than this, then the evolution will enter a forbidden regime. The abundance of the lightest particle (the DM candidate) decouples when the Boltzmann suppression factor due to the mass splitting becomes significant and the chain reaction departs from chemical equilibrium. After this point, the abundance of the heavier particles continues its Boltzmann suppression, $Y_{\rm heavy} \sim e^{-\Delta m/T}$. Consequently, the relic abundances of the heavier states are much smaller than that of the lightest state when in the forbidden regime. We leave a detailed study of this forbidden regime to future work.

The second important effect of the mass splitting is that the heavier states can quickly cascade decay to the lightest state (which is still very long lived). Depending on the size of the mass-splitting, the decays may be in- or out-of-equilibrium. If the decays are in-equilibrium, the calculation of the relic-abundance will mostly remain unchanged. If the decays are out of equilibrium, the heavier states will simply transfer their abundances to the lightest state, increasing the DM abundance by a factor $\mathcal{O}(N)$.

It is also reasonable to consider that the interaction strengths may vary across the chain. In this case, the Boltzmann equations are best solved by diagonalizing the system of  differential equations. The relic abundance is approximately given by the degenerate case, but with the cross-section replaced by the smallest eigenvalue cross-section in the chain.

\section{Phenomenology}

One of the main phenomenological signatures of the mechanism are decays of long-lived relics. Late time decaying DM ($\tau > 10^{10}$~yr) can potentially be probed for all masses. DM decays can change the CMB anisotropies and spectrum~\cite{Slatyer:2016qyl}, but a detailed study for very heavy dark matter has not been performed. Decays of super heavy DM can source ultra-high-energy cosmic rays (UHECR), with energies $\gtrsim 10^9$~GeV. UHECR can be observed in diffuse gamma ray satellites, such as FERMI-LAT~\cite{Ackermann:2014usa}, high energy neutrino experiments, such as IceCube~\cite{Abbasi:2011ji}, and dedicated UHECR observatories, such as  Auger~\cite{Abreu:2011zze}. For studies of the indirect detection of heavy decaying DM, see Refs.~\cite{Berezinsky:1997hy,Kuzmin:1998uv,Barbot:2002gt,Esmaili:2012us,Murase:2012xs,Feldstein:2013kka,Rott:2014kfa,Aloisio:2015lva,Kalashev:2016cre,Kalashev:2017ijd,Blanco:2018esa,Alcantara:2019sco}. Prospects for discovering UHECR are expected to improve with the proposed POEMMA mission~\cite{Olinto:2017xbi}.

 Depending on the parameters of the theory (especially the size of the mass splittings), some of the $\chi_i$ may decay at various cosmological epochs.  For instance, a component of the DM could dissociate light elements during Big Bang Nucleosynthesis~(see Ref.~\cite{Kawasaki:2017bqm} and references therein). Additionally, DM decay can lead to spectral distortions in the CMB~\cite{Hu:1993gc,Chluba:2011hw,Poulin:2016anj}, which would be probed by the proposed PiXiE experiment~\cite{Kogut:2011xw}. 
Partially decaying DM has also been shown to alleviate several cosmological tensions, such as the Hubble tension~\cite{Vattis:2019efj,Pandey:2019plg} and small scale structure puzzles~(see e.g.  Refs.~\cite{Wang:2014ina,Enqvist:2015ara}).  For further studies  of multicomponent dark matter with varying lifetimes, see Refs.~\cite{Dienes:2011ja,Dienes:2018yoq}. A detailed exploration of the phenomenology of our framework will be presented in upcoming work~\cite{future}.

\section{A toy model}
Having described the general mechanism, we now present a simple model realization. 
Consider the Lagrangian
\beq
{\cal L} \supset -\frac{1}{2}m_i \chi_i \chi_i - \delta m \chi_{i} \chi_{i+1} - y S \chi_i\chi_i - \mu S |H|^2 - \tilde{y} \chi_N L H,
\label{model1}
\eeq
where $\chi_i$ ($i=1,\cdots,N$) are left-handed Weyl fermions, $S$ is a SM singlet scalar field, $L$ is the lepton doublet, and $H$ is the SM Higgs.
The model is technically natural, since it respects a $Z_2^N$ symmetry that is broken by  $\delta m$. Any correction to off-diagonal masses must then depend on some power of $\delta m$. For simplicity, we take $\delta m$ to be flavor independent; relaxing the assumption does  not qualitatively alter the results.

The mass matrix is given by
\bea
M_{ij} = m_i \delta_{ij} + \delta m  ( \delta_{i,j+1} + \delta_{i,j-1}) .
\eea
The above structure is similar to a tight-binding model along a one-dimensional wire in a quantum mechanical system. 
It is well-known that the wavefunctions in such system are localized at each site~\cite{PhysRev.109.1492}. 
Assuming ${\Delta m \equiv  ( m_{\rm max} - m_{\rm min} ) \gg \delta m}$, the localization length is~\cite{IZRAILEV2012125,Craig:2017ppp}
\bea
\xi_{\rm loc}^{-1} \simeq \ln \frac{\Delta m}{2 \delta m }  - 1.
\eea  
Due to the localization, the mass eigenstate $\psi_i$ can be approximated as
\bea
\psi_i \sim \sum_j \chi_j \exp\left[ - \frac{|i-j|}{\xi_{\rm loc}} \right]\,.
\eea
In the mass basis,  the generated nearest neighbor interactions 
\beq
\mathcal{L} \supset y e^{-1/\xi_{\rm loc}} S \psi_{i} \psi_{i+1} \,,,.
\eeq
and non-nearest neighbor interactions are exponentially suppressed. 

The mixing can generate the direct decay ${\psi_1 \to  H + L}$, with width
\bea
\left( \frac{\Gamma_1}{\Gamma_N} \right)_{\rm mixing} \simeq 
 e^{ - 2N/\xi_{\rm loc} }\,.
 \label{decaymixing}
\eea
Taking $\Gamma_N = H(m)$, we find 
\beq
\xi_{\rm loc}^{-1} \gtrsim \frac{1}{N}\left( 62 +  \log \frac{m}{10^{10}~\rm GeV}\right)
\eeq
to ensure that $\tau \gtrsim 10^{27}$ sec. 
The ratio between $\Delta m$ and $\Delta m$ controls both the relic abundance and the lifetime of DM in this toy model. 
For the longevity of DM, a small localization length is preferred, while, for the correct relic abundance, the localization should be not so small as it could suppress the nearest neighbor interaction.
In Fig.~\ref{fig:minima_N}, we plot the minimum number of heavy particles $\chi_i$ needed for the stability and the observed abundance of dark matter. 

\section{Summary}
In this {\it Letter}, we presented a new freezeout mechanism for super heavy DM that freezes out with the SM within a standard cosmological history. The relic abundance is determined solely via its interactions with the SM. For a velocity-independent cross-section, we showed the DM mass could be as large as $m\sim10^{14}$~GeV within the perturbative unitary limit. In an upcoming paper we show how  velocity-dependent interactions, such as if the scattering was mediated by a light mediator, allows for DM to be as heavy as the Planck scale~\cite{future1}. 
\\

\begin{acknowledgments}
{\em Acknowledgments ---} We are grateful to Tim Cohen, Raffaele D'Agnolo, Jared Evans, Yuval Grossman, Kenny C.Y. Ng,  Jinhong Park, Josh Ruderman, Juri Smirnov, and the Weizmann Institute HEP lunch group
 for useful discussions. We especially thank Yonit Hochberg for useful discussions and comments on the manuscript.
	The work of EK is supported by the Israel Science Foundation (grant No.1111/17), by the Binational Science Foundation (grant No. 2016153) and by the I-CORE Program of the Planning Budgeting Committee (grant No. 1937/12).
\end{acknowledgments}

\bibliographystyle{h-physrev5}
\bibliography{ref}

\vspace{10 cm}
\clearpage
\newpage
\maketitle
\onecolumngrid
\begin{center}
{\bf \large Super heavy thermal dark matter}
\\
\vspace{0.3cm}
{ \it \large Supplemental Material}
\\ 
\vspace{0.3cm}
{Hyungjin Kim and Eric Kuflik}
\end{center}
\setcounter{equation}{0}
\setcounter{figure}{0}
\setcounter{table}{0}
\setcounter{section}{0}

In the supplemental material, we present an alternative approach to compute  the relic abundance of dark matter for the $N>2$ degenerate case. This method is easily generalized to non-degenerate masses and varying interaction strengths. 
In matrix form, the Boltzmann Eqs.~\eqref{last_eq} can be written as
\bea
\mathbf{Y} ' \approx \frac{\lambda}{x^2} \mathbf{M} 
\cdot \mathbf{Y} + \frac{\lambda}{x^2} \mathbf{n}_{N-1} Y_{\rm eq},
\label{app_boltzmann}
\eea
where $\mathbf{n}_{N-1} = (0,\cdots, 1)^T$, $\mathbf{Y} = (Y_1, \cdots, Y_{N-1})^T$, and we have replaced the abundance of $N^{\rm th}$ particle with equilibrium abundance, $Y_N = Y_{\rm eq}$.
The matrix $\mathbf{M}$ is given as
\bea
(\mathbf{M})_{ij} =   \alpha_i (  \delta_{i,j-1} - \delta_{ij} ) + (\delta_{i,j+1} - \delta_{ij} ),
\eea
where $\alpha_i = 1 - \delta_{i1}$ and $i, j = 1, \cdots , N-1$.

In order to solve the system, we need to diagonalize the matrix $\mathbf{M}$. 
Let $\widetilde{\mathbf{Y}}_k$ be an eigenvector of $\mathbf{M}$, satisfying the eigenvalue equation, $\mathbf{M} \cdot \widetilde{\mathbf{Y}}_k = \beta_k \widetilde{\mathbf{Y}}_k$. 
We find that the $k^{\rm th}$ eigenvector and eigenvalue are 
\bea
\beta_k &=& - 2 + 2\cos\theta_k,
\\
\widetilde{\mathbf{Y}}_k &=& \frac{2}{\sqrt{2N-1}} ( \sin[(N-1)\theta_k], \sin[(N-2)\theta_k],\cdots,\sin\theta_k ),
\eea
where $\theta_k = \pi \frac{2k - 1}{2N-1}$. 
The normalization was chosen such that $\widetilde{\mathbf{Y}}^T \widetilde{\mathbf{Y}} = \mathbf I$. 

The Boltzmann system admits Bloch wave-like eigenvectors, and, as a consequence, the eigenvalue shows a band structure.
If one interprets the index $i$ as a discretized one-dimensional wire, it is clear that each eigenvector represents a configuration of $\chi$ densities along this one-dimensional discrete wire, while the eigenvalue represents an effective strength of scattering processes. 
Eigenvectors with large $k$ corresponds to short wavelength fluctuations of $\chi$  densities along the wire, and has an ${\cal O}(1)$ eigenvalue, indicating that these fluctuations decay faster than modes with smaller eigenvalues.
On the other hand, the eigenvector $k=1$ has the smallest eigenvalue, $\beta_1 \simeq - \pi^2 /4 N^2$ . This corresponds to the longest wavelength fluctuation of the $\chi$  densities.
Whatever initial condition we choose, the distribution of $\chi$ abundances relaxes to the mode with the smallest eigenvalue, and  the final relic abundance of dark matter is determined by the freeze-out of this longest wavelength mode.

To compute the relic density of dark matter candidate, we need only consider the $k=1$ mode. The Boltzmann equation for this mode is
\bea
\widetilde{\mathbf{Y}}_1 ' \approx \frac{\lambda}{x^2} \beta_1 \widetilde{\mathbf{Y}}_1 +  \frac{\lambda}{x^2} \frac{2\sin\theta_1 }{\sqrt{2N-1}} Y_{\rm eq}.
\eea
Solving the equation, we find asymptotic value for $\widetilde{\mathbf{Y}}_1(\infty)$ as
\bea
\widetilde{\mathbf{Y}}_1(\infty) \simeq \left( \frac{2}{|\beta_1|} \frac{\sin\theta_1}{\sqrt{2N-1}} \right) Y_\infty(\lambda |\beta_1|)
\eea
where $Y_{\infty}$ is defined in Eq.~\eqref{relic_N2}. 
Rotating back to the original basis, we find
\bea
Y_i(\infty) 
\approx Y_{\infty}(\lambda|\beta_1|) \frac{4}{\pi}\cos\Big( \frac{\pi}{2} \frac{i-\frac12}{N-\frac12} \Big),
\label{Y_asymp}
\eea
which is consistent to the result obtained in continuum limit in the main text for $N\gg 1$.
In particular, the relic abundance depends on the combination $\lambda |\beta_1|$, which is nothing but the diffusion coefficient $D = \lambda \pi^2 / 4N^2$ in the continuum limit.

\end{document}